\documentclass[manuscript]{emulateapj}

\shorttitle{Enhancement of a light wall with external disturbances}
\shortauthors{Yang et al.}

\journalinfo{Accepted for publication in ApJL}

\submitted{Accepted for publication in ApJL}

\begin{document}

\title{Enhancement of a sunspot light wall with external disturbances}

\author{Shuhong Yang\altaffilmark{1,2}, Jun Zhang\altaffilmark{1,2}, and Robert Erd\'{e}lyi\altaffilmark{3}}

\altaffiltext{1}{Key Laboratory of Solar Activity, National
Astronomical Observatories, Chinese Academy of Sciences, Beijing
100012, China; shuhongyang@nao.cas.cn}

\altaffiltext{2}{College of Astronomy and Space Sciences, University
of Chinese Academy of Sciences, Beijing 100049, China}

\altaffiltext{3}{Solar Physics and Space Plasma Research Centre,
School of Mathematics and Statistics, University of Sheffield, Hicks
Building, Hounsfield Road, Sheffield S3 7RH, UK}

\begin{abstract}

Based on the \emph{Interface Region Imaging Spectrograph}
observations, we study the response of a solar sunspot light wall to
external disturbances. A flare occurrence near the light wall caused
material to erupt from the lower solar atmosphere into the corona.
Some material falls back to the solar surface, and hits the light
bridge (i.e., the base of the light wall), then sudden brightenings
appear at the wall base followed by the rise of wall top, leading to
an increase of the wall height. Once the brightness of the wall base
fades, the height of the light wall begins to decrease. Five hours
later, another nearby flare takes place, a bright channel is formed
that extends from the flare towards the light bridge. Although no
obvious material flow along the bright channel is found, some
ejected material is conjectured to reach the light bridge.
Subsequently, the wall base brightens and the wall height begins to
increase again. Once more, when the brightness of the wall base
decays, the wall top fluctuates to lower heights. We suggest, based
on the observed cases, that the interaction of falling material and
ejected flare material with the light wall results in the
brightenings of wall base and causes the height of the light wall to
increase. Our results reveal that the light wall can be not only
powered by the linkage of \emph{p}-mode from below the photosphere,
but may also be enhanced by external disturbances, such as falling
material.
\end{abstract}


\section{INTRODUCTION}

Above sunspot light bridges, some dynamic activities, such as
brightenings and surges, have been observed in the lower solar
atmosphere (Asai et al. 2001; Shimizu et al. 2009; Louis et al.
2014; Tian et al. 2014; Toriumi et al. 2015a; Robustini et al.
2016). By examining the Doppler characters or the intensity
variations in light bridges, a number of authors found evidence for
the existence of oscillations in light bridges (Sobotka et al. 2013;
Yuan et al. 2014; Yuan \& Walsh 2016). With the high tempo-spatial
resolution observations from the \emph{Interface Region Imaging
Spectrograph} (\emph{IRIS}; De Pontieu et al. 2014), Yang et al.
(2015) have found an ensemble of bright structures in the lower
atmosphere rooted in a light bridge of NOAA Active Region (AR)
12192, and named this ensemble \emph{light wall}. The most distinct
character of the light wall is the existence of a much brighter wall
top observed in \emph{IRIS} 1330 {\AA} images. The light wall showed
evidence of oscillations in height, appearing as successive upward
and downward motions of the wall top. This kind of oscillation,
i.e., joint rising and falling motion of neighbouring bright
structures, was also noted by Bharti (2015). We have interpreted
that these oscillations of the light wall are due to the leakage of
\emph{p}-modes from below the photosphere (Yang et al. 2015). As the
global resonant acoustic oscillations, the \emph{p}-modes are deemed
to leak sufficient energy to power shocks and thus drive upward
flows (De Pontieu et al. 2004; Klimchuk 2006). Hou et al. (2016a)
examined \emph{IRIS} observations between 2014 December and 2015
June, and found that most light walls originate above light bridges.

In the present paper, we report our findings that the light wall can
be significantly disturbed by external events, such as falling
material. Due to the external disturbances, the wall base brightens
and the wall height increases.

\section{OBSERVATIONS AND DATA ANALYSIS}

\begin{figure*}
\centering
\includegraphics
[bb=64 296 530 529, clip,angle=0,width=0.95\textwidth]{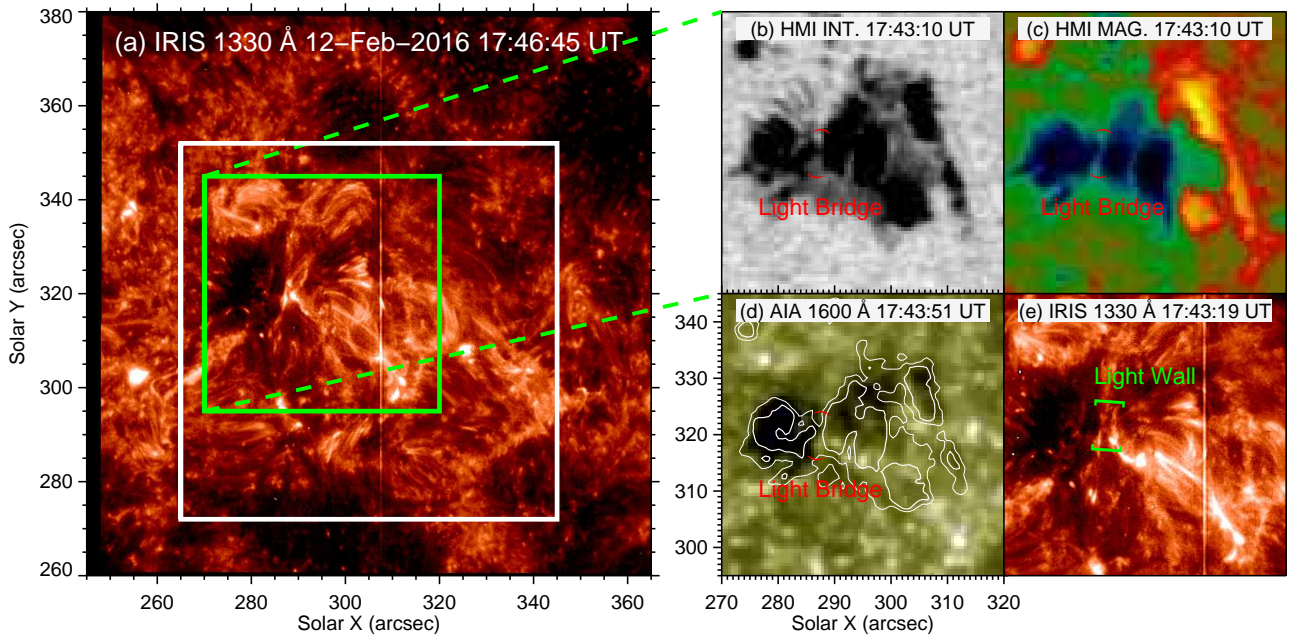}
\caption{(a) \emph{IRIS} SJI 1330 {\AA} image observed on 2016
February 12. The green square outlines the FOV of Figures 1(b-e),
and the white square outlines the FOV of Figure 2(a) and Figure 4.
(b-e) HMI intensity map, HMI magnetogram, AIA 1600 {\AA} image, and
\emph{IRIS} 1330 {\AA} image showing the light bridge (marked by the
red parentheses) and light wall (marked by the square brackets)
within AR 12497. The white curves are the contours of sunspots
identified in the HMI intensity map of panel (b). \label{fig1}}
\end{figure*}

We use two series of slit-jaw images (SJIs) observed by \emph{IRIS}
on 2016 February 12 in 1330 {\AA}. The 1330 {\AA} passband includes
a strong emission from the C II 1334/1335 {\AA} lines which are
formed in the chromosphere and lower transition region, and the
continuum emission from the upper photosphere and lower
chromosphere. The first series of SJIs were obtained from 17:23 UT
to 18:34 UT and the second one from 22:15 UT to 23:25 UT. Both of
the two series have a cadence of 10 s and a plate-scale of
0{\arcsec}.166. Their field-of-view (FOV) is 119{\arcsec} $\times$
119{\arcsec}, covering the main sunspot of NOAA AR 12497. The two
datasets have the same region (i.e., AR 12497) as target with
solar-rotation tracking. The spectral data were obtained on the
large coarse 4-step raster mode, and the slit crosses the path of
the falling material.

During the two \emph{IRIS} observational periods, we also obtained
simultaneous intensity maps and magnetograms from the Helioseismic
and Magnetic Imager (HMI; Scherrer et al. 2012; Schou et al. 2012)
on-board the \emph{Solar Dynamics Observatory} (\emph{SDO}; Pesnell
et al. 2012). The full-disk HMI data have a plate-scale of
0$\arcsec$.5 and a cadence of 45 s. In addition, we use the 1600
{\AA} images from the \emph{SDO} Atmospheric Imaging Assembly (AIA;
Lemen et al. 2012). The pixel size and cadence of the 1600 {\AA}
line are 0$\arcsec$.6 and 24 s, respectively. Each sequence of the
HMI and AIA data are calibrated to Level 1.5 with the standard
procedure \emph{aia\_prep.pro}, and differentially rotated to a
reference time. The reference times for the two sequences are 17:46
UT and 22:38 UT, respectively.

\section{RESULTS}

An overview of AR 12497 in \emph{IRIS} 1330 {\AA} is shown in Figure
1(a). In the HMI intensity map (panel (b)) at 17:43 UT, there is a
light bridge (outlined by the red parentheses) across the main
sunspot. In the corresponding HMI magnetogram, the magnetic field at
the light bridge is much weaker than the surrounding negative
fields. The light bridge can be identified both in the AIA 1600
{\AA} image (panel (d)) and \emph{IRIS} 1330 {\AA} image (panel
(e)). Moreover, a light wall, i.e., an ensemble of bright structures
rooted in the light bridge, is observed in the \emph{IRIS} 1330
{\AA} image, as denoted by the square brackets. The bright top of
the light wall is somewhat conspicuous. The light wall cannot be
identified in the AIA 1600 {\AA} image.

\subsection{Disturbance by the falling material}

\begin{figure*}
\centering
\includegraphics
[bb=91 209 513 632,clip,angle=0,width=0.6\textwidth]{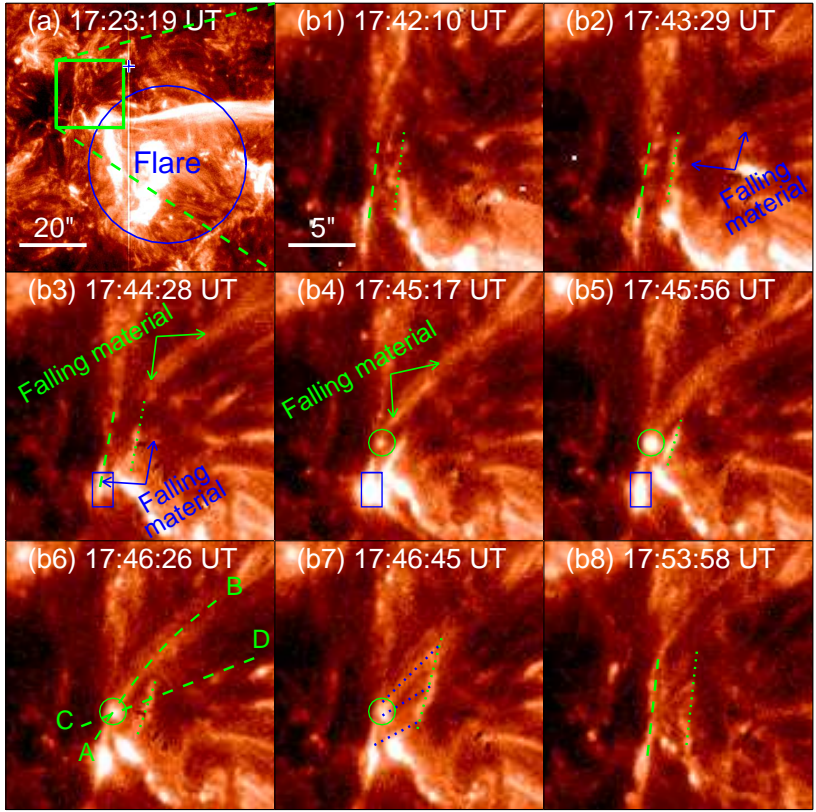}
\caption{(a) \emph{IRIS} 1330 {\AA} image displaying the occurrence
of a flare near the light wall (also see Movie 1). The square in
panel (a) marks the FOV of panels (b1-b8), and the blue circle
outlines the location where a flare occurred. The ``+" symbol marks
the position where the spectral profiles were analyzed. (b1-b8)
Sequence of \emph{IRIS} 1330 {\AA} images showing the falling
material (denoted by the arrows) and the evolution of the light
wall. The green dashed and dotted lines indicate the positions of
the wall base and wall top at different times. The blue rectangles
and green circles outline the landing points of the falling
material, also the brightening areas. The dashed curves ``A-B" and
``C-D" in panel (b6) mark the positions where the time-distance
plots are obtained in Figure 3. The blue dotted lines in panel (b7)
outline the bright threads connecting the top and base of the light
wall. \label{fig2}}
\end{figure*}

\begin{figure*}
\centering
\includegraphics
[bb=48 182 509 631,clip,angle=0,width=0.65\textwidth]{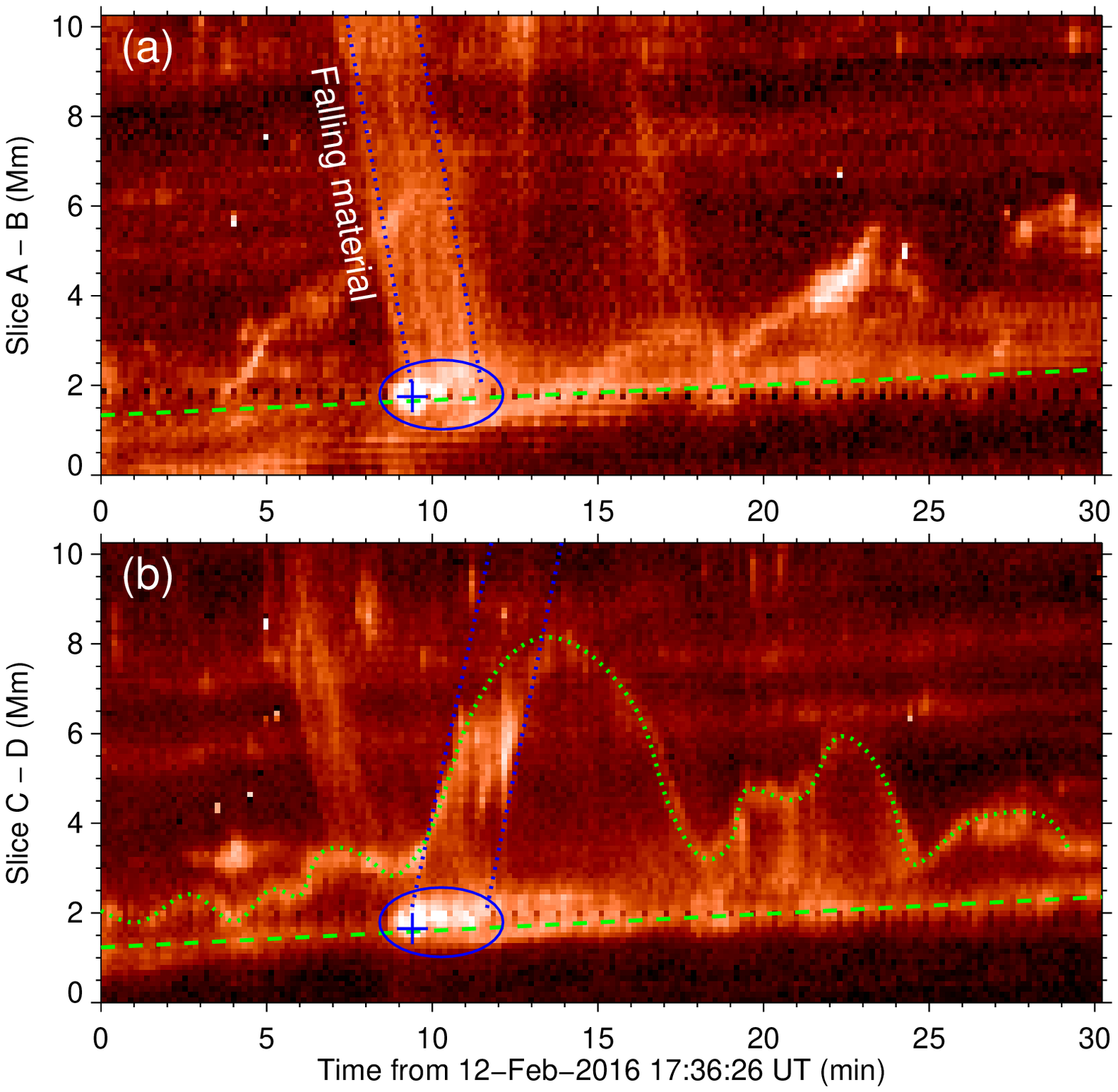}
\caption{(a-b) Time-distance plots along slice ``A-B" and ``C-D" of
Figure 2. The dashed lines mark the base of the light wall, and the
green dotted curve delineates the wall top. The blue dotted lines in
panel (a) outline the leading and trailing edges of the falling
material, and those in panel (b) outline the upward motion of the
wall top. The ``+" symbols mark the first landing point of the
falling material and brightening point. The ellipses outline the
recurrent brightenings at the base of the light wall. \label{fig3}}
\end{figure*}

In the south-west area near the light wall, a C6.8 flare occurred,
as outlined by the blue circle in Figure 2(a). After that, a great
deal of material was ejected into the corona (outlined by the blue
dotted circle in Movie 1), and some of it descended downwards to the
solar surface (denoted by the arrows in Movie 1). At 17:42:10 UT
(see panel (b1)), the wall base and wall top are marked by the green
dashed and dotted lines, respectively. Then, the falling material
(indicated by the blue arrows in panels (b2) and (b3)) fell towards
the base of the light wall. When the falling material reached the
wall base at 17:44:28 UT, the landing point (outlined by the blue
rectangle in panel (b3)) brightened. Meanwhile, more material as
denoted by the green arrows in panel (b3) was falling down. This
material hit the wall base at 17:45:17 UT, and there was also a
brightening at the landing point (outlined by the green circle in
panel (b4)). Then, 39 s later, the brightening became more obvious
(panel (b5)). Subsequently, the wall top began to rise (see panels
(b5)-(b7)). At 17:46:45 UT, the wall top reached at a higher
position (panel (b7)) compared to that of at 17:45:56 UT and
17:46:26 UT, respectively. Note that a large number of bright
threads connecting the top and base of the light wall can be
identified (marked by the blue dotted lines in panel (b7)),
revealing the fanning out structure of the light wall. This can be
used to explain why the wall with large increase in height is much
larger than the brightened region at the wall base. Several minutes
later, the top of the light wall began to fall back (see panel
(b8)). We found that the bottom part of the light wall started to
rise up prior to the main impact of the downflows. The reason may be
that material was already falling prior to the main impact, which
remains invisible at the \emph{IRIS} resolution and temperature
response.

To investigate the evolution of falling material and light wall, we
make two time-distance diagrams along slices ``A-B" and ``C-D"
(marked in Figure 2(b5)), respectively. The time-distance diagrams
are presented in Figure 3. The base of the light wall is marked by
the dashed line in each panel. The falling material appears as
bright structures, and the leading and trailing edges are outlined
by the two dotted lines in panel (a). When the falling material
reaches the wall base with the projected velocity of 71 km s$^{-1}$,
a range of brightenings appeared in the area outlined by the blue
ellipse. The ``+" symbol in panel (a) shows the first landing point
of the falling material and the first brightening point around
17:45:30 UT. In panel (b), the green dotted curve outlines the
variation of the bright wall top. When the first landing point
became brightened (marked by the ``+" symbol), the wall top rose
quickly with an average velocity of 60 km s$^{-1}$. Due to the
successive falling of material, several brightenings at the wall
base appeared and the wall top moved upward (between the two blue
dotted lines). The short streaks between the blue dotted lines are
the characteristic of material due to the side motion of bright
threads. The projected height of the light wall reached at the
maximum of about 6.5 Mm at 17:50 UT, which is much higher than its
height of 1.3 Mm just before the interaction with the falling
material. Due to the projection effect, the real height of the light
wall may be actually larger. Then, the height of the light wall
began to decrease. At 17:54 UT, the projected height was 1.2 Mm,
which is comparable to the one before the base brightenings. We
analyzed the spectral profiles of Si IV 1393.78 {\AA} in the path of
the falling material (marked by the ``+" symbol in Figure 2(a)). At
17:42:59 UT, the falling material was flowing across the slit, and
by applying the Gaussian fitting method, we found that the red-shift
relative to the background is about 29 km s$^{-1}$. Since the
projected speed is 71 km s$^{-1}$, the total speed of the downflow
is estimated to be 77 km s$^{-1}$.

\subsection{Disturbance by the flare}

\begin{figure*}
\centering
\includegraphics
[bb=89 277 515 563,clip,angle=0,width=0.85\textwidth]{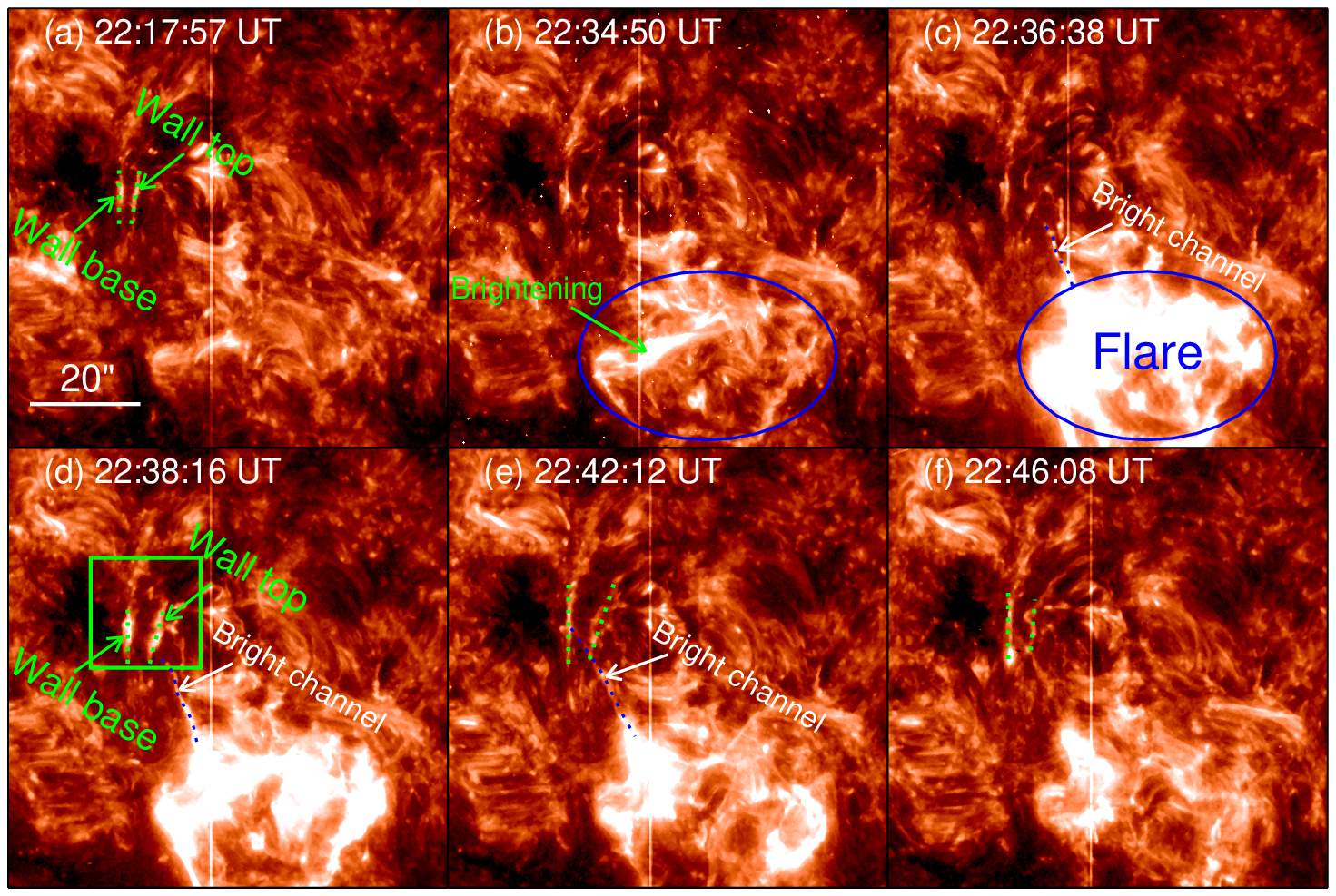}
\caption{Sequence of \emph{IRIS} 1330 {\AA} images displaying the
occurrence of another flare with a bright channel (blue dotted
curves) connecting the light bridge (see also Movie 2). The blue
ellipses outline the flare location. \label{fig4}}
\end{figure*}

\begin{figure*}
\centering
\includegraphics
[bb=85 196 495 641,clip,angle=0,width=0.56\textwidth]{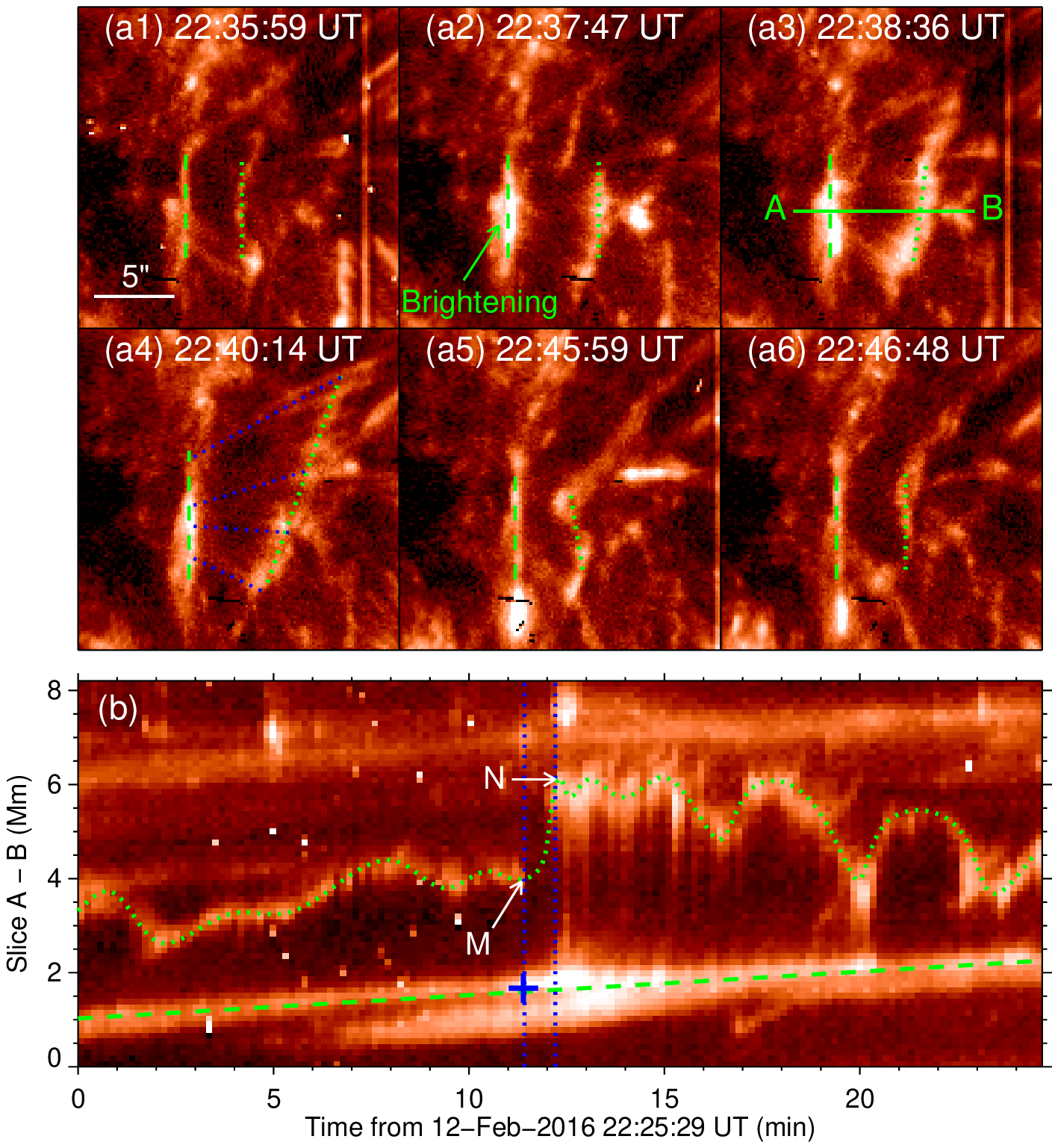}
\caption{(a1-a6) \emph{IRIS} 1330 {\AA} images showing the evolution
of the light wall impacted by the second flare. The green dashed and
dotted lines mark the base and top of the light wall, respectively.
The blue dotted lines in panel (a4) outline the fanning out bright
threads connecting the top and base of the light wall. (b)
Time-distance plot along slice ``A-B" (marked in panel (a3)) derived
from the 1330 {\AA} images. The dashed line outlines the wall base
and the dotted curve outlines the wall top, and the blue ``+" symbol
denotes the first brightening site. The left and right vertical
lines mark the start and peak times of the disturbance,
respectively. The arrows ``M" and ``N" denote the positions of the
wall top at the start and peak disturbing times, respectively.
\label{fig5}}
\end{figure*}

In the larger FOV, outlined by the white window in Figure 1(a), next
we study another flare and its effect on the light wall. At 22:17:57
UT, the base and top of the light wall can be well identified, and
there was no observable explosive event in the surrounding area (see
Figure 4(a)). At 22:34:50 UT, the region outlined by the blue
ellipse began to brighten (panel (b)), and then a C3.0 flare
occurred (panel (c)). We note that, during the flare process, a
bright channel which looks like a tail of the flare appeared, as
indicated by the white arrow in panel (c). The bright channel
extended from the flare towards the light wall. Then, the base of
the light wall brightened and the height of the light wall
increased. The bright channel was more evident at 22:38:16 UT (panel
(d)). Although the flare decayed, the bright channel linking the
flare and wall base remained clear for a while (panels (e)-(f)).

In order to study the response of the light wall during the flare,
we display in Figures 5(a1)-(a6) a sequence of \emph{IRIS} 1330
{\AA} images in a small FOV (outlined by the green square in Figure
4(d)). In Figures 5(a1)-(a6), the dashed lines and dotted lines
outline the wall base and wall top, respectively. Before the
disturbance by the flare, the projected height of the light wall is
about 3.6{\arcsec} at 22:35:59 UT (panel (a1)). At 22:37:47 UT, the
light wall was affected by the flare. Therefore the base of the
light wall was brightened as denoted by the arrow in panel (a2), and
the brightness of the wall top also increased, compared with that of
2 min earlier. Moreover, the wall top moved upwards quickly (panels
(a2)-(a4)). A set of bright threads (outlined by the blue dotted
lines in panel (b4)) connecting the wall base and wall top formed a
fan-shaped structure. After that, the emission of the wall base
decayed and the wall top moved downwards again (panels (a4)-(a6)),
reaching at a lower height. In order to study the evolution of the
light wall in more details, we make a time-distance plot along slice
``A-B" marked in panel (a3), and present it in panel (b). In the
time-distance plot, the green dashed line and dotted curve outline
the wall base and wall top, respectively. We may see that, before
22:37 UT (marked by the left vertical line), the top of the light
wall moved upward and downward with an average projected height of
about 2.5 Mm. At 22:37 UT, the height of the light wall itself was
about 2.5 Mm, as denoted by arrow ``M". Then, the wall base began to
brighten at 22:37 UT, as marked by the blue ``+" symbol.
Subsequently, the height of the light wall increased quickly in the
following 50 s (marked by the right vertical line) and became
approximately 4.5 Mm (denoted by arrow ``N") in height. In the next
3 minutes, the wall base brightened continually and the wall top
maintained its relatively high altitude. When the brightness of the
wall base decreased, the top of the light wall fluctuated to lower
heights. At last, the height of the light wall was comparable to
that before the disturbance.

\section{CONCLUSIONS AND DISCUSSION}

With \emph{IRIS} observations, we study the dynamical evolution and
response to external perturbations of a light wall. A flare occurred
near the light wall followed by material ejected into the corona
from the lower solar atmosphere. Some of the ejected material fell
back to the solar surface. When the falling material reached the
light bridge, i.e., the base of the light wall, sudden brightenings
appeared at the wall base and, most importantly, the wall top rose
quickly, performing increase of the wall height. Once the
brightenings of the wall base faded, the height of the light wall
began to decrease. When another nearby flare took place, a bright
channel was formed and extended from the flare towards the light
bridge. Then, the wall base brightened and the height of the light
wall began to increase again. Once again, when the brightness of the
wall base decreased, the wall top fluctuated to lower heights.

Falling material slides down along magnetic field lines (Reale et
al. 2013; Kleint et al. 2014; Innes et al. 2016; Jing et al. 2016).
We suggest that the falling material shown in Figure 2 causes the
brightenings of the wall base and the increase of the light wall
height. The proposed scenario is as follows: When the falling
material hits the light bridge, the kinetic energy is converted to
the thermal energy. Due to the local associated heating, the light
bridge brightens and the pressure of the plasma therein increases
concurrently. The magnetic configuration of light wall is thought to
be a group of magnetic field lines rooted in the light bridge (Hou
et al. 2016b). Thus, the material is lifted at a much higher speed
along magnetic field lines by the increased pressure at the bottom,
leading to the observational phenomenon that the top of the light
wall is powered to reach a greater height. A given magnetic flux
loop often guides a large number of falling elongated blobs (Antolin
et al. 2010; Antolin \& Rouppe van der Voort 2012). Indeed, we can
identify several blobs from the falling material (as shown in Movie
1), which appear as multi-trajectories marked between two dotted
lines in Figure 3(a). The falling material continuously crashes into
the wall base, leading to successive brightenings as outlined by the
ellipses in Figures 3(a)-(b). Consequently, multi-trajectories of
upward motion of the wall top are observed, as marked by two dotted
curves in Figure 3(b).

Applying the methods of Gilbert et al. (2013) and Innes et al.
(2016), the mass of falling material in Figure 3(a) is estimated to
be about 1.6$\times$10$^{12}$ g. Since the total speed of the
downflow is 77 km s$^{-1}$, then the kinetic energy is $\sim$
4.8$\times$10$^{25}$ erg. This energy can power the light wall
material to increase by 11 Mm in height. The observed height
increase of the light wall shown in Figure 3(b) is consistent with
this estimate. In addition, the rebound shock generated by the
impact of falling material could indeed play some role in leading to
the light wall's increase in height. Movie 1 and Figure 2 show that
the falling material and the light wall are two separated structures
in the corona, but they are joined at the footpoint (exactly very
close to each other). Thus the impact of the falling material seems
to affect a large region in the corona. This also implies that the
falling material does not need to fall along the same path as the
light wall to perturb it after the impact.

For the C3.0 flare, only a bright channel was observed. We suggest
that, although no obvious material flow along the bright channel was
found, some amount of ejected material may still reach the light
bridge. A somewhat similar process has been reported briefly by Yang
et al. (2014). In that study, on three homologous confined flares,
one remote region brightened when each flare occurred, which is also
deemed to be caused by ejected material. For the present study, when
the ejected particles along the bright channel impacted on the light
bridge, kinetic energy is likely be converted to thermal energy that
heated the wall base and powered the light wall. For this case, we
cannot fully exclude the possibility that the disturbance originates
from below, such as small-scale reconnection in the lower atmosphere
due to e.g. convection motion (Toriumi et al. 2015a, b). We note
that there was a brightening at the wall base at 22:33 UT (indicated
by the first arrow in Movie 2). This brightening only corresponds to
a small increase of the wall height, much smaller than that resulted
from the bright channel. Although it is difficult to determine the
exact cause of the brightening at 22:33 UT, we think it may be
caused by the small-scale reconnection due to the convection motion
(Toriumi et al. 2015a, b).

We conjecture that a light wall is a slab-like structure (maybe even
consisting of thin magnetic threads) embedded in vertically
stratified plasma. Oscillations of light wall have earlier been
interpreted to be powered by the leakage of \emph{p}-modes from
below the photosphere (Yang et al. 2015). Our new results reveal
that the light wall can also be enhanced by external disturbances,
such as falling material and an avalanche of particles caused by
nearby flares.

\acknowledgments {We thank the referee for the helpful suggestions
and constructive comments. The data are used courtesy of \emph{IRIS}
and \emph{SDO} science teams. This work is supported by the National
Natural Science Foundations of China (11673035, 11533008, 11373004,
11303049, 11221063), the Strategic Priority Research Program (No.
XDB09000000), the Youth Innovation Promotion Association of CAS
(2014043), the CAS Project KJCX2-EW-T07, and the Young Researcher
Grant of National Astronomical Observatories of CAS. RE is grateful
to STFC (UK) for the awarded Consolidated Grant, The Royal Society
for the support received in a number of mobility grants. He also
thanks the Chinese Academy of Sciences Presidents International
Fellowship Initiative, Grant No. 2016VMA045 for support received.}

{}


\begin{thebibliography}{}

\bibitem[Asai et al.(2001)]{2001ApJ...555L..65A} Asai, A., Ishii, T.~T., \& Kurokawa, H.\ 2001, \apjl, 555, L65

\bibitem[Antolin \& Rouppe van der Voort(2012)]{2012ApJ...745..152A} Antolin, P., \& Rouppe van der Voort, L.\ 2012, \apj, 745, 152

\bibitem[Antolin et al.(2010)]{2010ApJ...716..154A} Antolin, P., Shibata, K., \& Vissers, G.\ 2010, \apj, 716, 154

\bibitem[Bharti(2015)]{2015MNRAS.452L..16B} Bharti, L.\ 2015, \mnras, 452, L16

\bibitem[De Pontieu et al.(2004)]{2004Natur.430..536D} De Pontieu, B., Erd{\'e}lyi, R., \& James, S.~P.\ 2004, \nat, 430, 536

\bibitem[De Pontieu et al.(2014)]{2014SoPh..289.2733D} De Pontieu, B., Title, A.~M., Lemen, J.~R., et al.\ 2014, \solphys, 289, 2733

\bibitem[Gilbert et al.(2013)]{2013ApJ...776L..12G} Gilbert, H.~R., Inglis, A.~R., Mays, M.~L., et al.\ 2013, \apjl, 776, L12

\bibitem[Hou et al.(2016)]{2016A&A...589L...7H} Hou, Y.~J., Li, T., Yang, S.~H., \& Zhang, J.\ 2016a, \aap, 589, L7

\bibitem[Hou et al.(2016)]{2016ApJ...829L..29H} Hou, Y. J., Zhang, J., Li, T., Yang, S. H., et al.\ 2016b, \apjl, 829, L29

\bibitem[Innes et al.(2016)]{2016A&A...592A..17I} Innes, D.~E., Heinrich, P., Inhester, B., \& Guo, L. J.\ 2016, \aap, 592, A17

\bibitem[Jing et al.(2016)]{2016NatSR...624319J} Jing, J., Xu, Y., Cao, W., et al.\ 2016, Scientific Reports, 6, 24319

\bibitem[Kleint et al.(2014)]{2014ApJ...789L..42K} Kleint, L., Antolin, P., Tian, H., et al.\ 2014, \apjl, 789, L42

\bibitem[Klimchuk(2006)]{2006SoPh..234...41K} Klimchuk, J.~A.\ 2006, \solphys, 234, 41

\bibitem[Lemen et al.(2012)]{2012SoPh..275...17L} Lemen, J.~R., Title, A.~M., Akin, D.~J., et al.\ 2012, \solphys, 275, 17

\bibitem[Louis et al.(2014)]{2014A&A...567A..96L} Louis, R.~E., Beck, C., \& Ichimoto, K.\ 2014, \aap, 567, A96

\bibitem[Pesnell et al.(2012)]{2012SoPh..275....3P} Pesnell, W.~D., Thompson, B.~J., \& Chamberlin, P.~C.\ 2012, \solphys, 275, 3

\bibitem[Reale et al.(2013)]{2013Sci...341..251R} Reale, F., Orlando, S., Testa, P., et al.\ 2013, Science, 341, 251

\bibitem[Robustini et al.(2016)]{2016A&A...590A..57R} Robustini, C., Leenaarts, J., de la Cruz Rodriguez, J., \& Rouppe van der Voort, L.\ 2016, \aap, 590, A57

\bibitem[Scherrer et al.(2012)]{2012SoPh..275..207S} Scherrer, P.~H., Schou, J., Bush, R.~I., et al.\ 2012, \solphys, 275, 207

\bibitem[Schou et al.(2012)]{2012SoPh..275..229S} Schou, J., Scherrer, P.~H., Bush, R.~I., et al.\ 2012, \solphys, 275, 229

\bibitem[Shimizu et al.(2009)]{2009ApJ...696L..66S} Shimizu, T., Katsukawa, Y., Kubo, M., et al.\ 2009, \apjl, 696, L66

\bibitem[Sobotka et al.(2013)]{2013A&A...560A..84S} Sobotka, M., {\v S}vanda, M., Jur{\v c}{\'a}k, J., et al.\ 2013, \aap, 560, A84

\bibitem[Tian et al.(2014)]{2014ApJ...790L..29T} Tian, H., Kleint, L., Peter, H., et al.\ 2014, \apjl, 790, L29

\bibitem[Toriumi et al.(2015)]{2015ApJ...811..137T} Toriumi, S., Katsukawa, Y., \& Cheung, M.~C.~M.\ 2015a, \apj, 811, 137

\bibitem[Toriumi et al.(2015)]{2015ApJ...811..138T} Toriumi, S., Cheung, M.~C.~M., \& Katsukawa, Y.\ 2015b, \apj, 811, 138

\bibitem[Yang et al.(2014)]{2014ApJ...793L..28Y} Yang, S.~H., Zhang, J., \& Xiang, Y.~Y.\ 2014, \apjl, 793, L28

\bibitem[Yang et al.(2015)]{2015ApJ...804L..27Y} Yang, S.~H., Zhang, J., Jiang, F.~Y., \& Xiang, Y.~Y.\ 2015, \apjl, 804, L27

\bibitem[Yuan et al.(2014)]{2014ApJ...792...41Y} Yuan, D., Nakariakov, V.~M., Huang, Z., et al.\ 2014, \apj, 792, 41

\bibitem[Yuan \& Walsh(2016)]{2016A&A...594A.101Y} Yuan, D., \& Walsh, R.~W.\ 2016, \aap, 594, A101

\end{thebibliography}
\end{document}